\documentclass[aps,pra,reprint,amsmath,amssymb,showpacs,nobibnotes]{revtex4-1}

\usepackage{graphicx}

\usepackage{bm}     
\usepackage{hyperref} 
\usepackage{xcolor}
\hypersetup{
    colorlinks,
    linkcolor={blue!80!black},
    citecolor={blue!80!black},
    urlcolor={blue!80!black}
}
\usepackage{dcolumn} 
\usepackage{dsfont}    
\usepackage{color}
\usepackage{textcomp,gensymb} 
\begin{document}

\title{Dispersion cancellation in a quantum interferometer with independent single photons}

\author{Dong-Gil Im}
\email{eastgil7902@gmail.com}
\affiliation{Department of Physics, Pohang University of Science and Technology (POSTECH), Pohang, 37673, Korea}

\author{Yosep Kim}
\affiliation{Department of Physics, Pohang University of Science and Technology (POSTECH), Pohang, 37673, Korea}

\author{Yoon-Ho Kim}
\email{yoonho72@gmail.com}
\affiliation{Department of Physics, Pohang University of Science and Technology (POSTECH), Pohang, 37673, Korea}

\date{\today}

\begin{abstract}
A key technique to perform a proper quantum information processing is to get a high visibility quantum interference between independent single photons. One of the crucial elements that affects the quantum interference is a group velocity dispersion that occurs when the single photons pass through a dispersive medium. We theoretically and experimentally demonstrate that an effect of group velocity dispersion on the two-photon interference can be cancelled if two independent single photons experience the same amount of pulse broadening. This dispersion cancellation effect can be generalized to a multi-path linear interferometer with multiple independent single photons. As multi-path quantum interferometers are at the heart of quantum communication, photonic quantum computing, and boson sampling applications, our work should find wide applicability in quantum information science.  
\end{abstract}

\maketitle


Quantum interference effects are at the heart of various quantum information processing applications, including quantum communication \cite{Gisin07,Ursin07}, quantum teleportation \cite{Bennett93,Bouwmeester97,Kim01,wang2015quantum}, quantum imaging \cite{pittman95,strekalov95,dangelo04,Brida10}, quantum metrology \cite{Giovannetti11}, and quantum computing \cite{Kok07,yao12}. Therefore, in photonic quantum information, attaining high visibility quantum interference is often synonymous to demonstrating the desired quantum information tasks.  One of the crucial elements that affects the quantum interference is the group velocity dispersion that occurs when the single photons pass through a dispersive medium. Such dispersive medium can be an optical fiber to transport the single photons \cite{Korzh15} and other optical components in the interferometer. Especially, if the photon bandwidth is large, even for a thin medium, the dispersion effects should be considered and dealt with properly.

One of the simplest systems for the quantum interferometer is the two-photon interferometer \cite{Hong87} in which two indistinguishable single photons injected into each input mode of a 50/50 beam splitter become coalesced and found together in either one of the two output modes of the beam splitter due to the bosonic nature of photons. The dispersion effect on the two-photon interference with photon pairs generated via spontaneous parametric down-conversion (SPDC) process has been extensively studied for the fundamental aspects \cite{Chiao,Chiao_pra} and shown that a dispersive medium in one arm does not affect the two-photon interference if there is frequency anti-correlation between two entangled-photons. This effect of dispersion cancellation based on frequency anti-correlation between the entangled photons has been applied to particular applications such as quantum optical coherence tomography \cite{QOCT02,QOCT03} and quantum clock synchronization \cite{QCS}. It is worthwhile to note that frequency anti-correlation between the entangled photons also allows non-local dispersion cancellation in which the two photons never interact with each other at a beam splitter before detection \cite{baek09}.

\begin{figure}[b]
\centering
\includegraphics[width=3in]{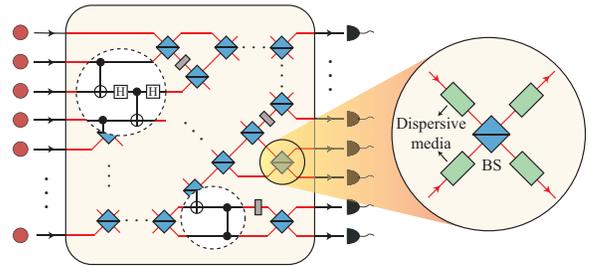}
\caption{
Schematic of a photonic quantum computing circuit. Linear optics are used for implementing quantum gates and all such optical elements introduce dispersion to the single photon pulses. 
}\label{fig:fig0}
\end{figure}

For quantum information processing applications, however, quantum interference with the independent single photons has broader applicability and, therefore, it is necessary to understand how to deal with the negative effects of dispersion in this scenario. In particular, an interesting question arises as to whether it would be possible to achieve dispersion cancellation in a multi-mode quantum interferometer with independent single photons, see Fig. \ref{fig:fig0}. 

In this paper, we theoretically and experimentally demonstrate that the effect of group velocity dispersion in a quantum interferometer  can be cancelled even with independent single photons at the input. Specifically, we show that Hong-Ou-Mandel-type two-photon interference with independent single photons at the input to the beam splitter exhibits cancellation of the group velocity dispersion effect if two independent single photons experience the same amount of pulse broadening before injected into a beam splitter. Each single photon pulse, prepared by heralding a photon of the photon pair born in the process of SPDC pumped by a femtosecond pulsed laser, becomes broadened as a result of propagation through a single-mode optical fiber \cite{baek08}. However, the two-photon interference does not exhibit any broadening, thereby exhibiting cancellation of group velocity dispersion in a quantum interferometer if the independent single photon pulse experiences the same amount of dispersion before the beam splitter. Moreover, our theoretical analyses show that this dispersion cancellation effect can be generalized to multi-path linear interferometer with multiple independent single photons which is used to implement a unitary transformation on single photons for the purpose of realizing quantum gates \cite{Hofmann02,Ralph02} for photonic quantum computation \cite{Knill01,Kok07,yao12,yosep19} and experimental boson sampling \cite{Tillmann13,Spring13,yosep20}.

\begin{figure}[t]
\centering
\includegraphics[width=3.4in]{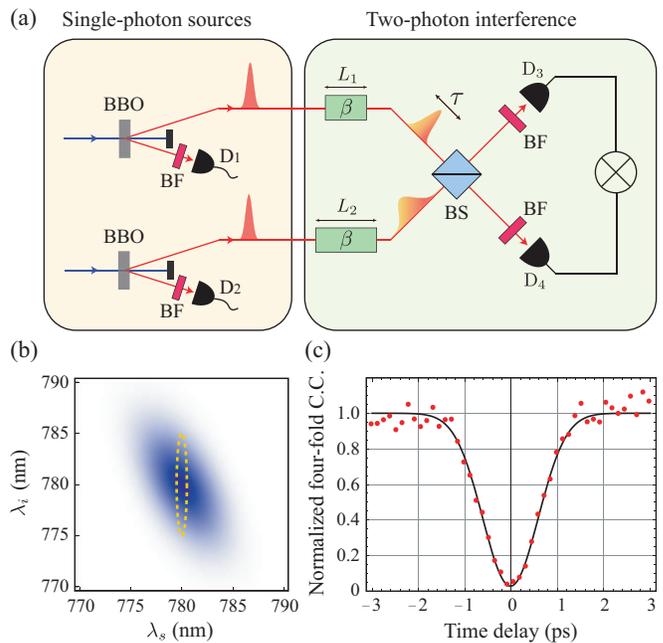}
\caption{
(a) Schematic of the experiment. Two  independent heralded single photons are prepared via the SPDC process and sent to a beam splitter (BS) after passing through the optical fibers of lengths  $L_1$ and $L_2$. The group velocity dispersion of the optical fiber is $\beta$.  (b) Calculated joint spectral intensity (JSI) of the SPDC photon pair. The yellow dashed ellipse represents the JSI when the heralded single photons are filtered by 1 nm FWHM bandpass filters (BF) while the heralding photons are filtered by 10 nm FWHM BFs. (c) With the heralded single photon bandwidth set at 1 nm FWHM, even with $L_1=L_2=6$ m, we observe high visibility quantum interference of 97\%. This is due to the fact that single photons of 1 nm bandwidth do not produce significant pulse broadening. The solid circles are the normalized four-fold coincidence count rates and the solid line represents Gaussian fits to the data points.
}\label{fig:fig1}
\end{figure}

The experimental setup is schematically shown in Fig.~\ref{fig:fig1}(a). Femtosecond laser pulses (80 MHz, 780 nm, temporal duration 140 fs) from a Ti:sapphire oscillator are frequency doubled and pump two 1-mm-thick type-II BBO crystals for the SPDC process. The non-collinear degenerate SPDC photons have the center wavelength of 780 nm and the heralding photons are detected at the detectors $\text{D}_1$ and $\text{D}_2$ after passing through 10 nm full width at half maximum (FWHM)  bandpass filters (BF). The two independent heralded single photons pass through the single-mode optical fibers  having the same group velocity dispersion $\beta$ and of lengths $L_1$ and $L_2$. The single photons are then injected into the two input ports of a 50/50 beam splitter for the Hong-Ou-Mandel type two-photon interference measurement. Finally, heralded single photons are detected at the detectors $\text{D}_3$ and $\text{D}_4$ after passing through BFs. We observed the two-photon interference pattern between the two heralded single photons by monitoring the four-fold coincidence count rates as a function of the relative optical time delay $\tau$ in the two arms. 

Since we make use of type-II SPDC for the photon pair generation, the signal and idler photons of SPDC have a specific frequency correlation or joint spectral intensity (JSI) as shown in Fig.~\ref{fig:fig1}(b). So, the 10 nm FWHM BFs placed in front of the heralding detectors $\text{D}_1$ and $\text{D}_2$ will cause tracing out of the idler photon spectral modes, resulting in spectrally mixed heralded single photon states for the signal photons. First, to ensure that the experimental setup is properly aligned, the BFs in front of the detectors $\text{D}_3$ and $\text{D}_4$ are set at 1 nm FWHM. The yellow dashed ellipse in Fig.~\ref{fig:fig1}(b) represents the JSI when the heralded single photons are filtered by 1 nm FWHM BFs while the heralding photons are filtered by 10 nm FWHM BFs. It is clear that the filtering process has removed frequency correlation between the signal and idler photons, allowing us to prepare spectrally pure heralded single-photon states.  Then, with the lengths of the optical fibers set at $L_1 = L_2=6$ m, we observe the four-fold coincidence count rates as a function of the optical delay $\tau$. The results of this experiment is shown in Fig.~\ref{fig:fig1}(c). With the heralded single photon bandwidth set at 1 nm FWHM, even with $L_1=L_2=6$ m, we observe high visibility quantum interference of 97\%. This is due to the fact that single photons of 1 nm bandwidth do not produce significant pulse broadening.
In order to estimate the background noise counts coming from the multiple pairs of SPDC photons which will lead to the erroneous heralding rates, we additionally measured the four-fold coincidence count rates by blocking one of the SPDC sources and then by blocking the other SPDC source.

To add noticeable dispersion effects to the heralded single photons for the optical fiber used in the experiment, it is necessary for us to change the lengths ($L_1$ and $L_2$) and the bandwidths of the single photons. Note that, for the fused silica core of the optical fiber, the group velocity dispersion parameter   $\beta$ at 780 nm is 37.802 $\text{fs}^2$/mm. First, the bandwidths of the heralded single photons are changed to 10 nm by using 10 nm FWHM BFs in front of detectors $\text{D}_3$ and $\text{D}_4$. Then the two-photon interference measurement is carried out by interfering two independent single photons by passing them through optical fibers of different lengths   ($L_1$ and $L_2$).

\begin{figure*}[!th]
\centering
\includegraphics[width=6.8in]{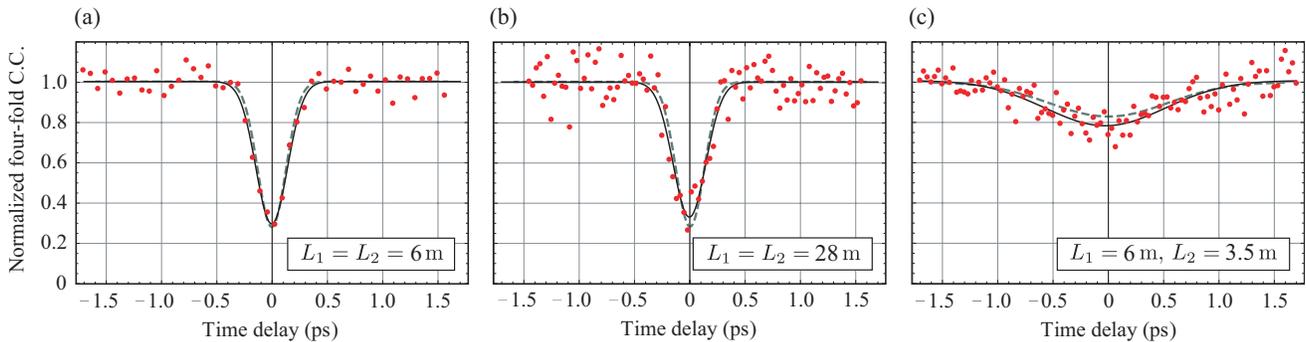}
\caption{
Experimental results confirming dispersion cancellation in a two-photon quantum interferometer with independent heralded single photons. The bandwidths of the heralded single photons are set at 10 nm. (a) When the optical fibers are of the same lengths, $L_1=L_2=6$ m, the group dispersion effects to the single photon is not observed in the two-photon quantum interference. (b) Even when the optical fiber lengths are increased to $L_1=L_2=28 \, \text{m}$,  two-photon quantum interference do not exhibit the group velocity dispersion effects. (c) When the lengths are different, $L_1=6 \, \text{m}, L_2=3.5 \, \text{m}$, the group dispersion effects are clearly seen with the broadening of the two-photon interference dip and reduction of the visibility. The solid circles are the normalized four-fold coincidence count rates and the solid lines represent Gaussian fits to the data points. The dashed lines show the results of the theoretical calculation of the two-photon interference. 
}\label{fig:fig2}
\end{figure*}

The experimental results confirming dispersion cancellation in a quantum interferometer with independent heralded single photons are shown in Fig.~\ref{fig:fig2}. First, the lengths of the optical fibers are set at $L_1=L_2=6$ m. In this condition, the 10 nm bandwidth single photon pulse is expected to broadened to 9.84 ps in time. However, the data in  Fig.~\ref{fig:fig2}(a) show clearly that the group velocity dispersion effect has been cancelled out: the FWHM width of the two-photon interference dip is 0.34 ps, significantly smaller than the expected pulse broadening. The two-photon visibility is  70.4\% in Fig.~\ref{fig:fig2}(a) which is smaller than that of Fig.~\ref{fig:fig1}(c), but this is unrelated to the group velocity dispersion effect. Rather, the reduced visibility is attributed to the frequency correlation nature of SPDC photons used in the experiment. While the increased single photon bandwidths of 10 nm do provide large enough bandwidths for noticeable pulse broadening, it also introduces unwanted spectral correlation between the heralding and the heralded photons, causing reduction of two-photon interference visibility \cite{zeilinger,Mosley08}.  Second, the lengths of the optical fibers are now set at  $L_1=L_2=28$ m and, in this condition,  the 10 nm bandwidth single photon pulse is expected to broadened to 45.9 ps in time. However, the experimental data shown in Fig.~\ref{fig:fig2}(b) clearly demonstrate that the dispersion effect has been cancelled out: the FWHM width of the two-photon interference dip is 0.35 ps, similarly to Fig.~\ref{fig:fig2}(a). The two-photon interference visibility is slightly reduced to 67.4\% from 70.4\% of Fig.~\ref{fig:fig2}(a) for the same reason mentioned above. Finally, when the lengths of the optical fibers are different, $L_1$ = 6 m, $L_2$ = 3.5 m, the effects of dispersion show up immediately, see Fig.~\ref{fig:fig2}(c): the FWHM width of the two-photon interference dip is broadened to 1.28 ps and the visibility has been significantly lowered to 22.9\%. Note that for the case of $L_1$ = 6 m, $L_2$ = 3.5 m, the optical fiber length difference between the two arms is compensated by the additional free space path so that the single photons from the both arms arrive simultaneously at the beam splitter.




The experimental results in Fig.~\ref{fig:fig2} make it clear that, when two independent single photons experience different amount of dispersion, the effects of dispersion appear at two-photon quantum interference. However, when the two photons experience the same amount of dispersion, the dispersion effect is cancelled out completely, exhibiting the same two-photon interference feature as if there had been no dispersion induced pulse broadening.  In order to describe the dispersion cancellation effect on the two-photon interference theoretically, we begin our study by writing the SPDC two-photon state in the Schmidt decomposed form which is useful for describing the heralded single photon state \cite{Mosley08},
\begin{equation}
\begin{split}\label{eq:01}
|\Psi^j \rangle& =  \int \int d\omega_{s} d\omega_{i} f^j(\omega_{s},\omega_{i}) e^{-i \theta_j(\omega_s)} \hat{a}_{s_j}^{\dagger}(\omega_{s}) \hat{a}_{i_j}^{\dagger}(\omega_{i}) |0\rangle 
\\& = \sum_{n}\sqrt{\lambda_n^j} \int d\omega_{s} \phi_n^j(\omega_{s}) e^{-i \theta_j(\omega_s)} \hat{a}_{s_j}^{\dagger}(\omega_{s}) 
\\& \times \int d\omega_{i} \psi_n^j(\omega_{i}) \hat{a}_{i_j}^{\dagger}(\omega_{i}) |0\rangle
\end{split}
\end{equation}
where $j=1(2)$ stands for the indices of the first (second) SPDC process and $\hat{a}_{s_j}^{\dagger}(\omega_{s})$ and $\hat{a}_{i_j}^{\dagger}(\omega_{i})$ are the creation operators for signal and idler photons with frequencies $\omega_{s}$ and $\omega_{i}$. Here, we assume that the signal photon is heralded and passes through the dispersive medium of length $L_j$, so we have introduced  the signal photon phase $\theta_j(\omega_s)$ to reflect the group velocity dispersion $\beta_j$ as $\theta_j(\omega_s)=\frac{1}{2} L_j \beta_j \omega_s^2$. 
The terms $\phi_n^j(\omega_{s})$ and the $\psi_n^j(\omega_{i})$ are the Schmidt modes defined by Schmidt decompositions of the joint spectral intensities, 
$f^j(\omega_{s},\omega_{i}) = \sum_n \sqrt{\lambda_n^j} \phi_n^j(\omega_{s}) \psi_n^j(\omega_{i})$ where $\lambda_n^j$ is the Schmidt eigenvalue. 
The heralding process will make the two-photon state into the reduced state in the density matrix form, 
\begin{equation}
\begin{split}\label{eq:02}
\rho_{s}^j=\text{Tr}_{i}[|\Psi^j \rangle \langle \Psi^j |]=\sum_{n}\lambda_n^j |\phi_n^j\rangle \langle \phi_n^j|
\end{split}
\end{equation}
where $|\phi_n^j\rangle = \int d\omega \phi_n^j(\omega) e^{-i \theta_j(\omega)} \hat{a}_{s_j}^{\dagger} (\omega)|0\rangle$. Then, the two independent heralded single photon states before the beam splitter is described as
\begin{equation}
\begin{split}\label{eq:03}
\rho^{in} = \rho_{s}^1 \otimes \rho_{s}^2 = \sum_{nn'}\lambda_n^1 \lambda_{n'}^2 (|\phi_n^1\rangle |\phi_{n'}^2\rangle)(\langle\phi_n^1| \langle\phi_{n'}^2|).
\end{split}
\end{equation}

\begin{figure}[t]
\centering
\includegraphics[width=3in]{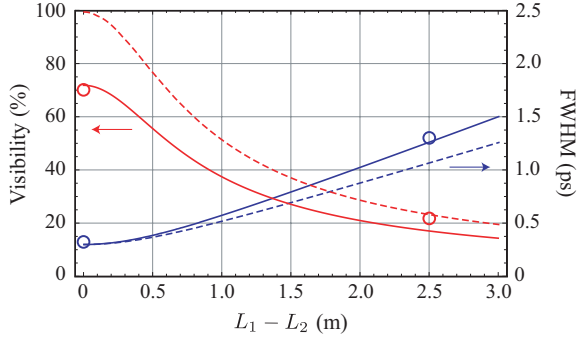}
\caption{
Theoretical visibility and FWHM widths of the two-photon interference dip between two independent heralded single photons as a function of the optical fiber lengths difference. The solid lines show the theoretical results using the experimental condition when the heralded single photons are in spectrally mixed states. The dashed lines show the theoretical results for the heralded single photons in the spectrally pure states having the same bandwidths. The solid circles show the experimental data.
}\label{fig:fig3}
\end{figure}

After applying the relative optical time delay $\tau$ and the beam splitter unitary transformation of the creation operators, the output state $\rho^{out}$ can be obtained. The coincidence probability P, between the two detectors $\text{D}_3$ and $\text{D}_4$ is given as
\begin{equation}
\begin{split}\label{eq:04}
& \text{P} =\text{Tr}[\rho^{out}\hat{P}_{\text{D}_3}\otimes \hat{P}_{\text{D}_4}]
\\& =\frac{1}{2} - \frac{1}{2}\sum_{nn'}\lambda_n^1\lambda_{n'}^2 \int d\omega \phi_n^{1}(\omega) \phi_{n'}^{2*}(\omega) e^{\frac{1}{2}i(\beta_1 L_1 - \beta_2 L_2) \omega^2} e^{i\omega \tau}
\\& \times \int d\omega' \phi_n^{1*}(\omega') \phi_{n'}^{2}(\omega') e^{-\frac{1}{2}i(\beta_1 L_1 - \beta_2 L_2) \omega'^2} e^{-i\omega' \tau}
\end{split}
\end{equation}
where $\hat{P}_{\text{D}_k}$ is the detection projector, $\hat{P}_{\text{D}_k} = \int d\omega \hat{b}_{s_k}^{\dagger}(\omega)|0\rangle \langle0| \hat{b}_{s_k}(\omega)$ on the output mode of the beam splitter $ \hat{b}_{s_k}^{\dagger}$.

If one postulates a pure state of the heralded single photon without changing the spectrum, it can be obtained by summing up the all Schmidt modes after multiplying the Schmidt eigenvalues, $\tilde{\phi}^{j} = \sum_n \lambda_n^j \phi_n^j (\omega_{s})$. By substituting the transformed Schmidt mode to the Eq.~(\ref{eq:02}), the coincidence probability can be obtained with the pure state which has Schmidt number of unity  satisfying $\lambda_n^1=\lambda_{n'}^2=1$, but the spectrum of photon remains same. In general, however, the Schmidt decomposition cannot be done analytically. So, we numerically calculated joint spectral intensity which is the product of the pump spectral profile and the phase matching function of the nonlinear crystal and computed its singular value decomposition, which is the matrix analogue of the Schmidt decomposition.
From the Eq.~(\ref{eq:04}), one can easily find that the dispersion effect is perfectly cancelled out when $\beta_1 L_1=\beta_2 L_2$ is satisfied. In our experiment, especially, we used the identical optical fibers for dispersive media, thus dispersion cancellation only depends on the optical fiber length difference between two optical fibers, $L_1-L_2$. 
By following the  Eq.~(\ref{eq:04}), we calculate the coincidence probability as a function of $\tau$ as shown in Fig.~\ref{fig:fig2} (dashed lines) and it is in good agreement with the experimental observations.
In addition, to examine the effects of dispersion, we calculated the visibility and FWHM width of the two-photon interference dip as a function of the optical fiber length difference as shown in Fig.~\ref{fig:fig3}.  Solid lines show the theoretical calculation following the experimental conditions where the heralded single photons are filtered by 10-nm FWHM bandwidth filters, so the single photon states are in the spectrally mixed state. As the optical fiber length difference is increased, the visibility is decreased and the FWHM bandwidth becomes broadened which results from the dispersion effect. If the heralded single photon states are in the spectrally pure state where the single photon has the same spectrum with the experimental condition, the dispersion is cancelled out and show the visibility of a unity when the optical fiber length difference is zero, but it shows the same tendency as the optical fiber length difference is increased (see the dashed lines in Fig.~\ref{fig:fig3}).

\begin{figure}[!t]
\centering
\includegraphics[width=2.9in]{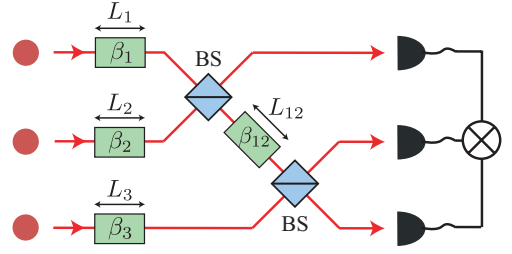}
\caption{
Schematic of a multi-path quantum interferometer with three single photons and four dispersive media. Under specific sets of conditions, full dispersion cancellation for three-photon quantum interference can be achieved.
 }\label{fig:fig5}
\end{figure}

It is also interesting to consider the dispersion cancellation for the multi-path linear interferometer with $N$ independent single photons. For instance, let us consider the scheme shown in Fig.~\ref{fig:fig5}.  Calculations show that the easiest way to achieve full dispersion cancellation for three-photon quantum interference is to satisfy the following conditions: (i) $\beta_1 L_1=\beta_2 L_2=\beta_3 L_3$ and $\beta_{12} L_{12}=0$ or (ii) $\beta_1 L_1=\beta_2 L_2$ and $\beta_1 L_1 + \beta_{12} L_{12} = \beta_3 L_3$. The key is to ensure that two photons interfering at a beam splitter to always undergo the same amount of dispersion. In this way, the dispersion cancellation scheme can be applied to $N$-photon  quantum circuits and networks


In conclusion, we have experimentally and theoretically investigated the dispersion effect on  quantum interference between two independent single photons and showed that it is possible to achieve full dispersion cancellation even with the independent single photons experiencing heavy group velocity dispersion as long as the photons experience the same amount of dispersion.   We have also shown that the dispersion cancellation in quantum interference can be generalized to N-photon quantum circuits and networks. Since multi-port quantum interferometers are at the heart of photonic quantum devices for quantum communication, quantum computing, and quantum simulation, we believe that our work would have wide applicability in quantum information science.

This work was supported in part by the National Research Foundation of Korea (Grant No. 2019R1A2C3004812), UG190028RD funded by Defense Acquisition Program Administration and Agency for Defense Development, and the MSIT of Korea under the ITRC support program (IITP-2020-0-01606).


\pagebreak

\end{document}